\documentclass[square]{ws-procs975x65}

\usepackage{array,booktabs}

\begin{document}

%
%
%

\title{SUPERCONFORMAL INDICES AND PARTITION FUNCTIONS FOR
  SUPERSYMMETRIC FIELD THEORIES}

\author{I. B. GAHRAMANOV$^{1,2}$ AND
  G. S. VARTANOV$^{1}$\footnote{Corresponding author. E-mail:
    grigory.vartanov@desy.de}}

\address{$^1$DESY Theory, Notkestr. 85, 22603 Hamburg, Germany }
\address{$^2$Institut f\"{u}r Physik, Humboldt-Universit\"{a}t zu Berlin,\\
  Newtonstrasse 15, 12489 Berlin, Germany}

\begin{abstract}
  Recently there was a substantial progress in understanding of
  supersymmetric theories (in particular, their BPS spectrum) in
  space-times of different dimensions due to the exact computation of
  superconformal indices and partition functions using localization
  method. Here we discuss a connection of $4d$ superconformal indices
  and $3d$ partition functions using a particular example of
  supersymmetric theories with matter in antisymmetric representation.
\end{abstract}


\bodymatter

\section{Introduction}
In a remarkable paper \cite{Dolan:2008qi} Dolan and Osborn recognized
the fact that the superconformal indices (SCIs) of $4d$ supersymmetric
gauge theories\cite{Kinney,Romelsberger1} are expressed in terms of
Spiridonov's elliptic hypergeometric integrals (EHI)\cite{S1}. This
observation provides currently the most rigorous mathematical
confirmation of $\mathcal{N}=1$ Seiberg electro-magnetic duality
\cite{Seiberg} through the equality of dual indices. The interrelation
between SCIs and EHIs was systematically studied\cite{SV1,SV2,V} and
there were found many new $\mathcal{N}=1$ physical dualities and also
conjectured new identities for EHIs. In particular, it was shown
\cite{Spiridonov:2012ww} that all 't Hooft anomaly matching conditions
for Seiberg dual theories can be derived from
$\mathrm{SL}(3,\mathbb{Z})$-modular transformation properties of the
kernels of dual indices. The theory of EHIs was applied also to a
description of the $S$-duality conjecture for $\mathcal{N}=2,4$
extended supersymmetric field theories \cite{Gadde:2011uv}. Several
modifications of SCIs have been considered recently such as the
inclusion of charge conjugation \cite{Zwiebel:2011wa}, indices on lens
spaces \cite{Benini:2011nc}, inclusion of surface operators
\cite{Nakayama:2011pa} or line operators
\cite{Dimofte:2011py,Gang:2012yr}.

By definition the SCI counts the BPS states protected by one
supersymmetry which can not be combined to form long multiplets. The
$\mathrm{SU}(2,2|1)$ space-time symmetry group of $\mathcal{N}=1$
superconformal algebra consists of $J_i, \overline{J}_i$, the
generators of two $\mathrm{SU}(2)$ subgroups forming the Lorentz group,
translations $P_\mu$, special conformal transformations $K_\mu$,
$\mu=1,2,3,4$, the dilatations $H$ and also the $U(1)_R$ generator
$R$. Apart from the bosonic generators there are supercharges
$Q_{\alpha},\overline{Q}_{\dot\alpha}$ and their superconformal
partners $S_{\alpha},\overline{S}_{\dot\alpha}$. Distinguishing a pair
of supercharges \cite{Romelsberger1}, for example, $Q=\overline{Q}_{1
}$ and $Q^{\dag}=-{\overline S}_{1}$, one has $\{Q,Q^{\dag}\}=
2{\mathcal H},\quad \mathcal{H}=H-2\overline{J}_3-3R/2, \label{susy}$
and then the superconformal index is defined by the matrix integral
\begin{equation}
  I(p,q,f_k) =  \text{Tr} \Big( (-1)^{\mathcal F}
  p^{\mathcal{R}/2+J_3}q^{\mathcal{R}/2-J_3}
  e^{\sum_{k}f_k F^k}e^{-\beta {\mathcal H}}\Big),
  \quad \mathcal{R}= R+2\overline{J}_3, \label{Ind}
\end{equation}
where ${\mathcal F}$ is the fermion number operator. Only zero modes
of $\mathcal H$ contribute to the trace because the commutation
relation for the supercharges is preserved by the operators used in
\eqref{Ind}. The chemical potentials $f_k$ are the group parameters of
the flavor symmetry group with the maximal torus generators $F^k$; $p$
and $q$ are group parameters for operators $\mathcal{R}/2\pm J_3$
commuting with $Q$ and $Q^{\dag}$.

According to the R\"omelsberger prescription \cite{Romelsberger1} for
${\cal N}=1$ superconformal theories one can write the full index via
a ``plethystic'' exponential
\begin{equation}
  I(p,q,\underline{y})  =  \int_{G_c} d \mu(g)\, \exp \bigg (
  \sum_{n=1}^{\infty} 
  \tfrac 1n \operatorname{ind}\big(p^n ,q^n, \underline{z}^n , \underline{y}^ n\big ) \bigg ),
\end{equation}
where $d \mu(g)$ is the $G_c$-invariant measure and single particle
states index
\begin{align} \nonumber
  \operatorname{ind}(p,q,\underline{z},\underline{y}) &= \frac{2pq - p
    - q}{(1-p)(1-q)} \chi_{adj}(\underline{z})
  \\
  \nonumber & \quad + \sum_j \frac{(p
    q)^{R_j/2}\chi_{R_F,j}(\underline{y})\chi_{R_G,j}(\underline{z}) -
    (pq)^{1-R_j/2}\chi_{{\bar R}_F,j}(\underline{y})\chi_{{\bar
        R}_G,j}(\underline{z})}{(1-p)(1-q)},
\end{align}
where the first term represents contributions of the gauge superfields
lying in the adjoint representation of the gauge group $G_c$. The sum
over $j$ corresponds to the contribution of chiral matter superfields
$\varphi_j$ transforming as the gauge group representations
$R_{{\scriptscriptstyle G,j}}$ and flavor symmetry group
representations $R_{{\scriptscriptstyle F,j}}$ with
$R_{\scriptscriptstyle j}$ being the field $R$-charges. The functions
$\chi_{{\scriptscriptstyle adj}}{\scriptstyle(\underline{z})}$,
$\chi_{{\scriptscriptstyle R_F,j}}{\scriptstyle(\underline{y})}$ and
$\chi_{{\scriptscriptstyle R_G,j}}{\scriptstyle(\underline{z})}$ are
the corresponding characters.

Let us consider the initial Seiberg duality\cite{Seiberg} for
SQCD. Namely, we take a $4d$ $\mathcal{N}=1$ SYM theory with
$G_c=\mathrm{SU}(N_c)$ gauge group and $N_f$ flavors with
$\mathrm{SU}(N_f)_l \times \mathrm{SU}(N_f)_r \times U(1)_B$ flavor
symmetry group. The original (electric) theory has $N_f$ left and
$N_f$ right quarks $Q$ and $\widetilde{Q}$ lying in fundamental and
anti-fundamental representation of the gauge group $\mathrm{SU}(N_c)$
and having $+1$ and $-1$ baryonic charges, $R=(N_f-N_c)/N_f$ is their
$R$-charge\footnote{This is the $R$-charge for the scalar component,
  the $R$-charge of the fermion component is $R-1$.}. The field
content of the described theory is summarized in the following table, where we have defined $\tilde{N_c}=N_f-N_c$:
\begin{center}
  \footnotesize
  \begin{tabular}{c !\quad c c c c c }
    \toprule
    & $\mathrm{SU}(N_c)$ & $\mathrm{SU}(N_f)_l$ & $\mathrm{SU}(N_f)_r$ & $U(1)_B$ & $U(1)_R$ \\
    \midrule
    $Q$ & $f$ & $f$ & 1 & 1 & ${\tilde{N_c}}/{N_f}$ \\
    $\widetilde{Q}$ & $\overline{f}$ & 1 & $\overline{f}$ & $-1$ & ${\tilde{N_c}}/{N_f}$ \\
    $V$ & $adj$ & 1 & 1 & 0 & 1 \\
    \bottomrule
  \end{tabular}
\end{center}

The corresponding SCI is given by the following elliptic
hypergeometric integral \cite{Dolan:2008qi}
\begin{align}\label{IE-seiberg}
  I_E = \kappa_{N_c} \int_{\mathbb{T}^{N_c-1}} \frac{\prod_{i=1}^{N_f}
    \prod_{j=1}^{N_c} \Gamma(s_i z_j,t^{-1}_i z^{-1}_j;p,q)} {\prod_{1
      \leq i < j \leq N_c} \Gamma(z_i z^{-1}_j,z_i^{-1} z_j;p,q)}
  \prod_{j=1}^{N_c-1} \frac{d z_j}{2 \pi \textup{i} z_j},
\end{align}
where $\prod_{j=1}^{N_c} z_j =1$. The balancing condition reads
$ST^{-1} = (pq)^{N_f-N_c}$ with $S =\prod_{i=1}^{N_f}s_i,$
$T=\prod_{i=1}^{N_f}t_i.$ The physical meaning of this condition is not completely  understood, one of the possible explanations is the independence of the index under marginal deformations.  We introduced the parameters $s_i$ and $t_i$
as
\begin{equation}
  s_i=(pq)^{R/2}vx_i, \qquad t_i=(pq)^{-R/2}vy_i,
  \label{ini_var}\end{equation}
where $x_i$, $y_i$ are chemical potentials for $\mathrm{SU}(N_f)_l$
and $\mathrm{SU}(N_f)_r$ groups satisfying the constraints
$\prod_{i=1}^{N_f}x_i=\prod_{i=1}^{N_f}y_i=1$, $v$ is the chemical
potential for $U(1)_B$-group, and
\begin{equation} \nonumber \kappa_{N_c} = \frac{(p;p)_{\infty}^{N_c-1}
    (q;q)_{\infty}^{N_c-1}}{N_c!}, \qquad
  (a;q)_\infty=\prod_{k=0}^\infty(1-aq^k).
\end{equation}
Here $\mathbb{T}$ denotes the unit circle with positive orientation
and we use conventions $\Gamma(a,b;p,q):=\Gamma(a;p,q)\Gamma(b;p,q),$
$\Gamma(az^{\pm1};p,q):=\Gamma(az;p,q)\Gamma(az^{-1};p,q)$, where
\begin{equation}\label{ellg}
  \Gamma(z;p,q)= \prod_{i,j=0}^\infty
  \frac{1-z^{-1}p^{i+1}q^{j+1}}{1-zp^iq^j}, \quad |p|, |q|<1,
\end{equation}
is the elliptic gamma function.

The dual (magnetic) theory is described by a $4d$ $\mathcal{N}=1$ SYM
theory with the gauge group $\widetilde{G}_c=\mathrm{SU}(N_f-N_c)$
sharing the same flavor symmetry\cite{Seiberg}. Here one has dual
quarks $q$ and $\widetilde{q}$ lying in the
fundamental/anti-fundamental representation of~$\widetilde{G}_c$,
which have $U(1)_B$-charges $N_c/(N_f-N_c)$, $-N_c/(N_f-N_c)$ and the
$R$-charge $N_c/N_f$, and additional mesons -- singlets of
$\widetilde{G}_c$ lying in the fundamental representation of
$\mathrm{SU}(N_f)_l$ and anti-fundamental representation of
$\mathrm{SU}(N_f)_r$ ($M_i^j=Q_i\widetilde{Q}^j,
i,j=1,\ldots,N_f$). It is convenient to collect again all field data
in one table:
\begin{center}
  \footnotesize
  \begin{tabular}{c !\quad c c c c c }
    \toprule
    & $\mathrm{SU}(\tilde{N}_c)$ & $\mathrm{SU}(N_f)_l$ & $\mathrm{SU}(N_f)_r$ & $U(1)_B$ & $U(1)_R$ \\
    \midrule
    $M$ & 1 & $f$ & $\overline{f}$ & 0 & $2\tilde{N}_c/N_f$ \\
    $q$ & $f$ & $\overline{f}$ & 1 & $N_c/{\tilde{N}_c}$ & ${N_c}/{N_f}$ \\
    $\widetilde{q}$ & $\overline{f}$ & 1 & $f$ & $-{N_c}/{\tilde{N}_c}$ & ${N_c}/{N_f}$ \\
    $V$ & $adj$ & 1 & 1 & 0 & 1 \\
    \bottomrule
  \end{tabular}
\end{center}
These two SQCD-type theories are dual to each other in their infrared
fixed points when the magnetic theory has the tree level
superpotential \cite{Seiberg}, $W = M_i^j q^i
\widetilde{q}_j$. The SCI of the magnetic theory is
\begin{align}\label{IM-seiberg}
  I_M ={}& \kappa_{N_{\widetilde{N}_c}}q \prod_{{\scriptscriptstyle 1 \leq
      i,j \leq N_f}}
  \Gamma({\scriptstyle s_i t^{-1}_j;p,q})
  \nonumber
  \\
  &\times\int_{\mathbb{T}^{*}} \prod_{j=1}^{\widetilde{N}_c-1} \frac{d \widetilde{z}_j}{2 \pi \textup{i} \widetilde{z}_j}
  \frac{\prod_{i=1}^{N_f} \prod_{j=1}^{\widetilde{N}_c}
    \Gamma({\scriptstyle S^{1/\widetilde{N}_c} s_i^{-1} \widetilde{z}_j,T^{-1/\widetilde{N}_c} t_i \widetilde{z}_j^{-1};p,q})}{\prod_{1 \leq i < j \leq \widetilde{N}_c}
    \Gamma({\scriptstyle \widetilde{z}_i \widetilde{z}_j^{-1},\widetilde{z}_i^{-1}\widetilde{z}_j;p,q})}  ,
\end{align}
where $\widetilde{N}_c=N_f-N_c$, $\prod_{j=1}^{\widetilde{N}_c}
\widetilde{z}_j = 1$ and
$\mathbb{T}^{*}=\mathbb{T}^{\widetilde{N}_c-1}$.

As discovered by Dolan and Osborn\cite{Dolan:2008qi}, the equality of
SCIs $I_E=I_M$ coincides with a mathematical identity established for
$N=2, N_f=3, 4$\cite{S1} and for arbitrary parameters\cite{Rains}.

\section{The anti-symmetric tensor matter field}

Recently the connection of $4d$ SCIs and $3d$ PFs was found \cite{DSV,Gadde:2011ia,Imamura:2011uw}
and the simplest example of SQCD type theory with $\mathrm{SP}(2N)$ gauge group
was considered. Here we would like to consider more complicated cases
with additional matter content. We start from the duality for $4d$
supersymmetric theory with the $\mathrm{SP}(2N)$ group introduced by
Intriligator \cite{Intriligator2}. The matter content of electric and
magnetic theories are given below in tables, respectively:
\begin{center}
  \footnotesize
  \begin{tabular}{c !\quad ccc}
    \toprule
    & $\mathrm{SP}(2N)$ & $\mathrm{SU}(2N_f)$ & $U(1)_R$ \\  
    \midrule
    $Q$ & $f$ & $f$ & $2r=1-\frac{2(N+K)}{(K+1)N_f}$ \\
    $X$ & $T_A$ & 1 & $2s=\frac{2}{K+1}$ \\
    \bottomrule
  \end{tabular}
\end{center}


\begin{center}
  \footnotesize
  \begin{tabular}{c !\quad ccc}
    \toprule
    & $\mathrm{SP}(2\widetilde{N})$ & $\mathrm{SU}(2N_f)$ & $U(1)_R$ 
    \\  
    \midrule
    $q$ & $f$ & $\overline{f}$ & $2\widetilde{r}=1-\frac{2(\widetilde{N} +K)}{(K+1)N_f}$
    \\
    $Y$ & $T_A$ & 1 & $2s=\frac{2}{K+1}$ 
    \\
    $M_j$  & 1 & $T_A$ & $2r_j=2\frac{K+j}{K+1}- 4 \frac{\widetilde{N}
      +K}{(K+1)N_f}$ 
    \\
    \bottomrule
  \end{tabular}
\end{center}
\noindent where $j=1, \ldots, K,$ and $\widetilde{N}  =  K(N_f-2)-N$,
$K=1,2,\ldots$

Defining $U=(pq)^s=(pq)^{\frac{1}{K+1}}$, we find the following
indices for these theories\cite{SV2}
\begin{align}\label{KSsp1_1}
  I_E ={}& \frac{(p;p)_{\infty}^{N} (q;q)_{\infty}^{N} }{2^N N!}
  \Gamma(U;p,q)^{N-1}
  \\
  \nonumber & \times \int_{\mathbb{T}^N} \prod_{1 \leq i < j \leq N}
  \frac{\Gamma(U z_i^{\pm 1} z_j^{\pm 1};p,q)}{\Gamma(z_i^{\pm 1}
    z_j^{\pm 1};p,q)} \prod_{j=1}^{N} \frac{\prod_{i=1}^{2N_f}
    \Gamma(s_i z_j^{\pm 1};p,q)} {\Gamma(z_j^{\pm
      2};p,q)}\prod_{j=1}^{N} \frac{d z_j}{2 \pi \textup{i} z_j},
  \displaybreak[3]
  \\
  \label{KSsp1_2}
  I_M ={}& \frac{(p;p)_{\infty}^{\widetilde{N}}
    (q;q)_{\infty}^{\widetilde{N}} }{2^{\widetilde{N}} \widetilde{N}!}
  \Gamma(U;p,q)^{\widetilde{N}-1} \prod_{l=1}^K \prod_{1 \leq i < j
    \leq 2N_f} \Gamma(U^{l-1} s_i s_j;p,q)
  \\
  \nonumber & \times \int_{\mathbb{T}^{\widetilde{N}}} \prod_{1
    \leq i < j \leq \widetilde{N}} \frac{\Gamma(U z_i^{\pm 1} z_j^{\pm
      1};p,q)}{\Gamma(z_i^{\pm 1} z_j^{\pm 1};p,q)}
  \prod_{j=1}^{\widetilde{N}} \frac{\prod_{i=1}^{2N_f} \Gamma(U
    s_i^{-1} z_j^{ \pm 1};p,q)} {\Gamma(z_j^{\pm
      2};p,q)}\prod_{j=1}^{\widetilde{N}}\frac{d z_j}{2 \pi \textup{i}
    z_j},
\end{align}
where the balancing condition reads $U^{2(N+K)}\prod_{i=1}^{2N_f}s_i=
(pq)^{N_f}.$

Using the asymptotic formula for the elliptic gamma function
\begin{align}
  \Gamma(e^{2 \pi \textup{i} r z};e^{2 \pi \textup{i} r \omega_1},
  e^{2 \pi \textup{i} r \omega_2})
  {\ \mathrel{\mathop{=}\limits_{r \rightarrow 0}}} e^{-\pi
    \textup{i}(2z-(\omega_1+\omega_2))/24r\omega_1\omega_2}
  \gamma^{(2)}(z;\omega_1,\omega_2),
\end{align}
where $\gamma^{(2)}(z)$ is a hyperbolic gamma function, one can
proceed with the reduction of SCIs for a dual pair presented
above. Let us reparameterize the variables in \eqref{KSsp1_1} and
\eqref{KSsp1_2} in the following way $p=e^{2 \pi \textup{i} v
  \omega_1}, \ q=e^{2 \pi \textup{i} v \omega_2}, \ s_i = e^{2 \pi
  \textup{i} v \alpha_i}, \ z_j=e^{2 \pi \textup{i} v u_j},$,
$i=1,\ldots,2N_f, \; j=1,\ldots,N$. Then after limit $v \rightarrow 0$, which assumes $pq\rightarrow1$,
one gets\footnote{Omitting the same divergent coefficients $\exp
  \left( \frac{-2 \pi i (-1 + K - 6 K N - 4 N^2)
      (\omega_1+\omega_2)}{24 v \omega_1 \omega_2 (1 + K)}\right)$.}
\begin{align}
  \label{KSsp1_hyp_1}
  I^{red}_E = {} & \frac{1}{{\scriptstyle 2^N N!}}
  \gamma({\scriptstyle
    \frac{\omega_1+\omega_2}{K+1}})^{{\scriptscriptstyle N-1}}
  \int_{-\textup{i} \infty}^{\textup{i} \infty}
  \prod_{{\scriptscriptstyle 1 \leq i < j \leq N}}
  \frac{\gamma({\scriptstyle\frac{\omega_1+\omega_2}{K+1} \pm u_i \pm
      u_j})}{\gamma({\scriptstyle\pm u_i \pm u_j})} \prod_{j=1}^{N}
  \frac{\prod_{i=1}^{2N_f} \gamma({\scriptstyle\alpha_i \pm u_j})}
  {\gamma({\scriptstyle\pm 2 u_j})} \frac{d u_j}{{\scriptstyle
      \textup{i} \sqrt{\omega_1 \omega_2}}}, \displaybreak[3]
  \\
  \label{KSsp1_hyp_2}
  I^{red}_M ={}& \frac{1}{2^{\widetilde{N}} \widetilde{N}!}
  \gamma({\scriptstyle
    \frac{\omega_1+\omega_2}{K+1}})^{\widetilde{N}-1} \prod_{l=1}^K
  \prod_{1 \leq i < j \leq 2N_f} \gamma({\scriptstyle (l-1)
    \frac{\omega_1+\omega_2}{K+1} + \alpha_i + \alpha_j})
  \\
  \nonumber & \times \int^{\textup{i} \infty}_{-\textup{i} \infty}
  \prod_{1 \leq i < j \leq \widetilde{N}} \frac{\gamma({\scriptstyle
      \frac{\omega_1+\omega_2}{K+1} \pm u_i \pm
      u_j})}{\gamma({\scriptstyle \pm u_i \pm u_j})}
  \prod_{j=1}^{\widetilde{N}} \frac{\prod_{i=1}^{2N_f} \gamma(
    {\scriptstyle\frac{\omega_1+\omega_2}{K+1} - \alpha_i \pm u_j})}
  {\gamma({\scriptstyle \pm 2 u_j})} \prod_{j=1}^{\widetilde{N}}
  \frac{d u_j}{\textup{i} \sqrt{\omega_1 \omega_2}},
\end{align}
where the balancing condition reads {\small $(\omega_1+\omega_2)
  \frac{2(N+K)}{(K+1)} + \sum_{i=1}^{2N_f} \alpha_i= N_f
  (\omega_1+\omega_2)$}. Above and in the rest of the paper, we use
the following notation $\gamma({\scriptstyle z})\equiv
\gamma^{(2)}({\scriptstyle z;\omega_1, \omega_2})$ and conventions
$\gamma({\scriptstyle a,b}) \equiv \gamma({\scriptstyle a})
\gamma({\scriptstyle b}),$ $ \gamma({\scriptstyle a\pm u}) \equiv
\gamma({\scriptstyle a+u}) \gamma({\scriptstyle a-u}).$


\subsection{Dualities for $\mathrm{SP}(2N)$ gauge group}

Let us consider now $\alpha_{2N_f} = \xi_1 + a S, \ \alpha_{2N_f-1} =
\xi_2 - a S$ and take the limit $S \rightarrow \infty$, then
$I_E^{red}$ and $I_M^{red}$ become
\begin{align}
  \label{KSsp1_Z_1}
   Z_E ={}& \frac{1}{{\scriptstyle 2^N N!}}  \gamma({\scriptstyle
    \frac{\omega_1+\omega_2}{K+1}})^{N-1} \int_{-\textup{i}
    \infty}^{\textup{i} \infty} \prod_{{\scriptscriptstyle 1 \leq i <
      j \leq N}}
  \frac{\gamma({\scriptstyle\frac{\omega_1+\omega_2}{K+1} \pm u_i \pm
      u_j})}{\gamma({\scriptstyle \pm u_i \pm u_j})} \prod_{j=1}^{N}
  \frac{\prod_{i=1}^{2(N_f-1)} \gamma({\scriptstyle \alpha_i \pm
      u_j})} {\gamma({\scriptstyle \pm 2 u_j})} \frac{d
    u_j}{{\scriptstyle \textup{i} \sqrt{\omega_1 \omega_2}}}
  \\
  \label{KSsp1_Z_2}
   Z_M ={}& \frac{1}{{\scriptstyle 2^{\widetilde{N}} \widetilde{N}!}}
  \gamma({\scriptstyle
    \frac{\omega_1+\omega_2}{K+1}})^{\widetilde{N}-1} \prod_{l=1}^K
  \gamma\Big({\scriptstyle(\omega_1+\omega_2)\big(
    N_f-\frac{2N+2K-l+1}{K+1}\big) - \sum_{i=1}^{2(N_f-1)} \alpha_i}
  \Big)
  \\
  \nonumber &  \times \prod_{l=1}^K \prod_{1 \leq i < j \leq
    2(N_f-1)} \gamma\big({\scriptstyle (l-1)
    \frac{\omega_1+\omega_2}{K+1} + \alpha_i + \alpha_j}\big)
  \\
  \nonumber &  \times \int_{-\textup{i} \infty}^{\textup{i}
    \infty} \prod_{1 \leq i < j \leq \widetilde{N}}
  \frac{\gamma({\scriptstyle \frac{\omega_1+\omega_2}{K+1} \pm u_i \pm
      u_j})}{\gamma({\scriptstyle \pm u_i \pm u_j})}
  \prod_{j=1}^{\widetilde{N}} \frac{\prod_{i=1}^{2(N_f-1)}
    \gamma({\scriptstyle \frac{\omega_1+\omega_2}{K+1} - \alpha_i \pm
      u_j})} {\gamma({\scriptstyle \pm 2 u_j})}
  \prod_{j=1}^{\widetilde{N}} \frac{d u_j}{\textup{i}
    {\scriptstyle\sqrt{\omega_1 \omega_2}}}.
\end{align}
To obtain these expressions we used the inversion relation
$\gamma({\scriptstyle z,\omega_1+\omega_2-z}) = 1$ and the asymptotic
formulas
\begin{align} \label{asympt} \lim_{u \rightarrow \infty} e^{\frac{\pi
      \textup{i}}{2} B_{2,2}(u;\omega_1,\omega_2)}
  \gamma({\scriptstyle u})
  & =  1, \text{ \ \ for } \text{arg }\omega_1 < \text{arg } u < \text{arg }\omega_2 + \pi, \nonumber \\
  \lim_{u \rightarrow \infty}e^{-\frac{\pi \textup{i}}{2}
    B_{2,2}(u;\omega_1,\omega_2)} \gamma({\scriptstyle u}) & = 1,
  \text{ \ \ for } \text{arg } \omega_1 - \pi < \text{arg } u <
  \text{arg }\omega_2,
\end{align}
where $B_{2,2}(u;\mathbf{\omega})$ is the second order Bernoulli
polynomial,
\begin{equation}
  B_{2,2}(u;\mathbf{\omega}) =
  \frac{u^2}{\omega_1\omega_2} - \frac{u}{\omega_1} -
  \frac{u}{\omega_2} + \frac{\omega_1}{6\omega_2} +
  \frac{\omega_2}{6\omega_1} + \frac 12.
\end{equation}
Note here, that the balancing condition is absent. Expressions
\eqref{KSsp1_Z_1} and \eqref{KSsp1_Z_2} reproduces the partition
functions of $3d$ $\mathcal{N}=2$ supersymemtric field theories
\cite{Jafferis:2010un,Hama:2010av}. Equality of \eqref{KSsp1_Z_1} and
\eqref{KSsp1_Z_2} gives us the duality for the $3d$ ${\cal N}=2$ SYM
theories with the matter content presented in the below tables:
\begin{center}
  \footnotesize
  \begin{tabular}{c !\quad cccc}
    \toprule
    & $\mathrm{SP}(2N)$ & $\mathrm{SU}(2(N_f-1))$ & $U(1)_A$ & $U(1)_R$ \\
    \midrule
    $Q$ & $f$ & $f$ & 1 & $1$ \\
    $X$ & $T_A$ & $1$ & 0 & ${2}/({K+1})$ \\
    \bottomrule
  \end{tabular}
\end{center}
\begin{center}
  \footnotesize
  \begin{tabular}{c !\quad cccc}
    \toprule
    & $\mathrm{SP}(2(K(N_f-2)-N))$ & $\mathrm{SU}(2(N_f-1))$ &
    $U(1)_A$ & $U(1)_R$ ($j=1,\ldots,K$)
    \\
    \midrule
    $q$ & $f$ & $\overline{f}$ & $-1$ & $\frac{3-K}{K+1}$ 
    \\
    $x$ & $T_A$ & $1$ & 0 & $\frac{2}{K+1}$ 
    \\
    $Y_j$ & $1$ & $1$ & $-2(N_f-1)$ & $4N_f-\frac{4N+6K-2j+4}{K+1}$ 
    \\
    $M_j$ & $1$ & $T_A$ & $2$ & $1 + 2 \frac{j-1}{K+1}$ 
    \\
    \bottomrule
  \end{tabular}
\end{center}

One can proceed with the reduction of flavors and take the limit
$\alpha_{2N_f-2} \rightarrow \infty$ after which one gets the equality
for PFs of the Chern-Simons Theory (CS) theories. Let us set $N_f
\rightarrow N_f-2$, then the electric theory is $3d$ $\mathcal{N}=2$
CS (!) theory with $k=1/2$ and the magnetic theory is $3d$
$\mathcal{N}=2$ CS theory with $k=-1/2$.

Now one can proceed further in integrating out the quarks by taking
further limits $s_i \rightarrow \infty$. As the result one gets the
extension for Kutasov-Schwimmer duality in three dimensions: the
electric theory is $3d$ $\mathcal{N}=2$ CS theory with
$\mathrm{SP}(2N)$ gauge group and level $k$ (such as $N_f+k$ is even),
$N_f$ quarks (which can be also odd~\cite{Willett:2011gp}), a chiral
superfield $X$ in adjoint representation, and the magnetic theory is
$3d$ $\mathcal{N}=2$ CS theory with $\mathrm{SP}(K(N_f+2(k-1))-2N)$
gauge group and level $-k$, $N_f$ quarks,a chiral superfield in
adjoint representation of the gauge group, mesons in $T_A$
representation of $\mathrm{SU}(N_f)$ global symmetry group.

\subsection{Dualities for $U(N)$ gauge groups}

We now consider different limit for the equality between
\eqref{KSsp1_hyp_1} and \eqref{KSsp1_hyp_2}. Let us reparameterize the
parameters in the following way $\alpha_i \rightarrow \alpha_i + \mu,
\ \alpha_{i+N_f} \rightarrow \alpha_{i+N_f} - \mu, \ i = 1, \ldots,
N_f$ and take the limit $\mu \rightarrow \infty$ after which one gets (for $K=1$ it coincides with the expression by Bult \cite{Bult})
\begin{align} \label{KSu1_hyp_1} I^{red,U(N)}_E ={}& \frac{1}{N!}
  \gamma({\scriptstyle \frac{\omega_1+\omega_2}{K+1}})^{N-1}
  \int_{-\textup{i} \infty}^{\textup{i} \infty} \prod_{j=1}^{N}
  \frac{d u_j}{\textup{i} \sqrt{\omega_1 \omega_2}}
  \\
  \nonumber & \times \prod_{1 \leq i < j \leq N}
  \frac{\gamma({\scriptstyle \frac{\omega_1+\omega_2}{K+1} \pm
      (u_i-u_j)})}{\gamma({\scriptstyle \pm (u_i-u_j)})}
  \prod_{j=1}^{N} \prod_{i=1}^{N_f} \gamma({\scriptstyle\alpha_i +
    u_j, \alpha_{i+N_f} - u_j})
\end{align} 
and
\begin{align} \label{KSu1_hyp_2} & I^{red,U(N)}_M =
  \frac{1}{\widetilde{N}!}
  \gamma({\scriptstyle\frac{\omega_1+\omega_2}{K+1}})^{\widetilde{N}-1}
  \prod_{l=1}^K \prod_{i,j=1}^{N_f} \gamma({\scriptstyle (l-1)
    \frac{\omega_1+\omega_2}{K+1} + \alpha_i + \alpha_{j+N_f}})
  \int_{-\textup{i} \infty}^{\textup{i} \infty}
  \prod_{j=1}^{\widetilde{N}} \frac{d
    u_j}{\textup{i} \sqrt{\omega_1 \omega_2}} \nonumber \\
  & \qquad \times \prod_{1 \leq i < j \leq \widetilde{N}}
  \frac{\gamma({\scriptstyle \frac{\omega_1+\omega_2}{K+1} \pm
      (u_i-u_j)})}{\gamma({\scriptstyle\pm (u_i-u_j)})}
  \prod_{j=1}^{\widetilde{N}} \prod_{i=1}^{N_f}
  \gamma({\scriptstyle\frac{\omega_1+\omega_2}{K+1} - \alpha_i - u_j,
    \frac{\omega_1+\omega_2}{K+1} - \alpha_{i+N_f} + u_j}),
\end{align}
where the balancing condition reads $(\omega_1+\omega_2)
2\frac{N+K}{K+1} + \sum_{i=1}^{N_f} (\alpha_i + \alpha_{i+N_f}) = N_f
(\omega_1+\nobreak\omega_2).$ Now considering the following reparametrization
\begin{equation}
  \alpha_{N_f-1} = \xi_1 + \mu, \quad \alpha_{N_f} = \xi_3 - \nu, \quad \alpha_{2N_f-1} = \xi_2 - \mu, \quad \alpha_{2N_f} = \xi_4 + \nu
\end{equation}
with the following limit $\mu \rightarrow \infty$ and $\nu \rightarrow
\infty$ one can obtain the PFs. Since verifying dualities for
$U(N)$ gauge groups is quite similar procedure to which was done
above, we only comment briefly on matter content of these
theories. More detailed explanations can be found in the original
papers\cite{DSV}.

The electric theory is $3d$ $\mathcal{N}=2$ SYM theory with the matter
content presented in the below table:
\begin{center}
  \footnotesize
  \begin{tabular}{c !\quad ccccc}
    \toprule
    & $U(N)$ & $\mathrm{SU}(N_f-2)$  & $\mathrm{SU}(N_f-2)$ &
    $U(1)_A$ & $U(1)_R$ 
    \\
    \midrule
    $Q$ & $f$ & $f$ & 1 & 1 & $1/2$ 
    \\
    $\widetilde{Q}$ & $\overline{f}$ & 1 & $f$ & 1 & $ 1/2$ 
    \\
    $X$ & $adj$ & $1$ & 1 & 0 & ${2}/{(K+1)}$ 
    \\
    \bottomrule
  \end{tabular}
\end{center}
The magnetic theory is $3d$ $\mathcal{N}=2$ SYM theory with the
matter content presented in the below table:
\begin{center}
  \footnotesize
  \begin{tabular}{c !\quad cccccc}
    \toprule
    & $U(\tilde N)$ & $\mathrm{SU}(N_f-2)$ &
    $\mathrm{SU}(N_f-2)$ & $U(1)$ & $U(1)_A$ & $U(1)_R$ 
    \\
    \midrule
    $q$ & $f$ & $\overline{f}$ & 1 & 0 & $-1$ & $\frac{3-K}{2(K+1)}$ 
    \\
    $\widetilde{q}$ & $\overline{f}$ & 1 & $\overline{f}$ & 0 & $-1$ &
    $\frac{3-K}{2(K+1)}$ 
    \\
    $x$ & $adj$ & $1$ & 1 & 0 & 0 & $\frac{2}{K+1}$ 
    \\
    $Y^{(1,2)}_j$ & $1$ & $1$ & 1 & $\pm1$ & $-(N_f-2)$ & $R_{Y_j}$ 
    \\
    $M_j$ & $1$ & $f$ & $f$ & 0 & $2$ & $1 + 2 \frac{j-1}{K+1}$ 
    \\
    \bottomrule
  \end{tabular}
\end{center}
where $\tilde N = K(N_f-2)-N$, $j=1,\ldots,K$ and
$R_{Y_j}=(N_f-2(N-j))/(K+1)$. For $K=1$ as in four-dimensional case
the duality goes to the Aharony duality\cite{Aharony:1997gp}.

The above duality is the duality between two $3d$ $\mathcal{N}=2$ SYM
(not CS) theories, namely between $3d$ $\mathcal{N}=2$ SYM (electric)
theory with $U(N)$ gauge group, $N_f$ quarks in fundamental and
anti-fundamental representation, a chiral superfield in adjoint
representation and $3d$ $\mathcal{N}=2$ SYM magnetic theory with
$U(KN_f-N)$ gauge group $N_f$ quarks in fundamental and
anti-fundamental representation, a chiral superfield in adjoint
representation, mesons in $(f,f)$ representation of $\mathrm{SU}(N_f-2) \times
\mathrm{SU}(N_f-2)$ global symmetry groups, chiral superfields $Y^{(1,2)}_j,
j=1,\ldots,K$. By doing similar reductions one can end up with the duality for CS theories, which coincide with the duality suggested by Niarchos \cite{Niarchos:2008jb}.

One can obtain CS theories by the following integrating out the matter
fields. For example, integrating out a pair of quarks by taking the
limit $\alpha_{N_f-3}, \alpha_{2N_f-3} \rightarrow \infty$ one gets
the following equality of PFs of the $3d$ $\mathcal{N}=2$ CS electric
theory with CS level equals to 1 and the $3d$ $\mathcal{N}=2$ CS
magnetic theory with CS level equals to $-1$ which coincides with the
results of Kapustin et al\cite{Kapustin:2011vz}.

\section*{Acknowledgments}
We would like to thank the organizers of the ``International congress
on mathematical physics'', Aalborg, 6--11 August 2012. IG wishes to
thank Jan Plefka for comments that helped us improve the text. GV
thanks V.P. Spiridonov for many fruitful discussions.


\begin{thebibliography}{99}

\bibitem{Dolan:2008qi} F.~A.~Dolan and H.~Osborn,
  Nucl. Phys.  B {\bf 818} (2009), 137--178.

\bibitem{Kinney} J. Kinney, J. M. Maldacena, S. Minwalla and S. Raju,
  Commun. Math. Phys. {\bf 275} (2007), 209--254.

\bibitem{Romelsberger1} C. R\"omelsberger,
  Nucl. Phys. B {\bf 747} (2006), 329--353;
  arXiv:0707.3702 [hep-th].

\bibitem{S1} V. P. Spiridonov,
  Uspekhi Mat. Nauk {\bf 56} (1) (2001), 181--182 (Russian
  Math. Surveys {\bf 56} (1) (2001), 185--186);
  Algebra i Analiz {\bf 15} (6) (2003), 161--215 (St.  Petersburg
  Math. J. {\bf 15} (6) (2004), 929--967);
  Uspekhi Mat. Nauk {\bf 63} (3) (2008), 3--72 (Russian Math. Surveys
  {\bf 63} (3) (2008), 405--472).

\bibitem{Seiberg} N. Seiberg,
  Nucl. Phys. {\bf B435} (1995), 129--146, {\tt hep-th/9411149}.

\bibitem{SV1} V. P. Spiridonov and G. S. Vartanov,
  Nucl. Phys. {\bf B824} (2010), 192--216;
  Phys. Rev. Lett. {\bf 105} (2010) 061603;
  Lett.\ Math.\ Phys.\ {\bf 100} (2012) 97;
  arXiv:1107.5788 [hep-th].
 
\bibitem{SV2} V.~P.~Spiridonov and G.~S.~Vartanov,
  Commun. Math. Phys. {\bf 304} (2011), 797--874.

\bibitem{V} G.~S.~Vartanov,
  Phys. Lett.  {\bf B696} (2011), 288--290.


\bibitem{Spiridonov:2012ww} V.~P.~Spiridonov and G.~S.~Vartanov, JHEP
  {\bf 1206} (2012) 016.


\bibitem{Gadde:2011uv} A.~Gadde, L.~Rastelli, S.~S.~Razamat and
  W.~Yan,
  {\tt arXiv:1110.3740 [hep-th]}; A. Gadde, E. Pomoni, L. Rastelli and
  S. S. Razamat,
  JHEP {\bf 03} (2010) 032.


\bibitem{Zwiebel:2011wa} B.~I.~Zwiebel,
  JHEP {\bf 1201} (2012) 116.


\bibitem{Benini:2011nc} F.~Benini, T.~Nishioka and M.~Yamazaki,
  Phys.\ Rev.\ D {\bf 86} (2012) 065015.

\bibitem{Nakayama:2011pa} Y.~Nakayama,
  JHEP {\bf 1108} (2011) 084.

\bibitem{Dimofte:2011py} T.~Dimofte, D.~Gaiotto and S.~Gukov,
  {\tt arXiv:1112.5179 [hep-th]}.


\bibitem{Gang:2012yr} D.~Gang, E.~Koh and K.~Lee,
  JHEP {\bf 1205} (2012) 007.

\bibitem{Rains} E. M. Rains,
  Ann. of Math. {\bf 171} (2010), 169--243.

\bibitem{DSV} F.~A.~H.~Dolan, V.~P.~Spiridonov and G.~S.~Vartanov,
  Phys.\ Lett.\ B {\bf 704} (2011) 234.

\bibitem{Gadde:2011ia}
  A.~Gadde and W.~Yan,
  JHEP {\bf 1212} (2012) 003.
  
\bibitem{Imamura:2011uw}
  Y.~Imamura,
  JHEP {\bf 1109} (2011) 133.

\bibitem{Intriligator2} K. Intriligator,
  Nucl. Phys. {\bf B448} (1995), 187--198, {\tt hep-th/9505051}.

\bibitem{Hama:2010av} N.~Hama, K.~Hosomichi and S.~Lee,
  JHEP {\bf 1103} (2011) 127;
  JHEP {\bf 1105} (2011) 014.
  
\bibitem{Jafferis:2010un} D.~L.~Jafferis,
  JHEP {\bf 1205} (2012) 159.
  
\bibitem{Willett:2011gp} B.~Willett, I.~Yaakov,
  {\tt arXiv:1104.0487 [hep-th]}.
  
\bibitem{Bult} F. van de Bult, 
PhD thesis, University of Amsterdam, 2007.

\bibitem{Aharony:1997gp} O.~Aharony,
  Phys.\ Lett.\ B {\bf 404} (1997) 71.

\bibitem{Niarchos:2008jb} V.~Niarchos,
  JHEP {\bf 0811} (2008), 001.

\bibitem{Kapustin:2011vz} A.~Kapustin, H.~Kim and J.~Park,
  JHEP {\bf 1112} (2011) 087.
  


\end{thebibliography}
\end{document}